\documentclass[10pt,conference,letterpaper]{IEEEtran}

\usepackage{amssymb}
\usepackage{amsmath}
\usepackage{balance}

\usepackage[colorlinks=true, linkcolor=blue, citecolor=blue, urlcolor=blue]{hyperref}

\usepackage{graphicx}
\usepackage{booktabs}
\usepackage{multirow}
\usepackage{tabularx}
\usepackage{threeparttable}
\usepackage{url}

\newcommand{\aaaa}{\overline{a}}
\newcommand{\na}{n_{\overline{a}}}

\newcommand{\vca}{\textup{\textbf{\textrm{a}}}}
\newcommand{\vcaaaa}{\overline{\textup{\textbf{\textrm{a}}}}}
\newcommand{\vcp}{\textup{\textbf{\textrm{p}}}}
\newcommand{\vcm}{\textup{\textbf{\textrm{m}}}}

\usepackage{fancyhdr} 
\fancypagestyle{firststyle}
{
   \fancyhf{}
   \fancyfoot[C]{\scriptsize{\copyright~IFIP 2020. This is the author's version of the work. It is posted here by permission of IFIP for your personal use. Not for redistribution. The definite version was published in: Proceedings of the IFIP Networking Conference (Networking 2020), Paris (online), IEEE, pp.~815--820, \url{https://ieeexplore.ieee.org/document/9142814}.}}
}

\begin{document}

\title{Measuring Basic Load-Balancing and Fail-Over Setups for Email Delivery via DNS MX Records}
\author{
\IEEEauthorblockN{Jukka Ruohonen}
\IEEEauthorblockA{University of Turku, Finland \\
Email: juanruo@utu.fi}
}

\maketitle

\begin{abstract}
The domain name system (DNS) has long provided means to assure basic load-balancing and fail-over (BLBFO) for email delivery. A~traditional method uses multiple mail exchanger (MX) records to distribute the load across multiple email servers. Round-robin DNS is the common alternative to this MX-based balancing. Despite the classical nature of these two solutions, neither one has received particular attention in Internet measurement research. To patch this gap, this paper examines BLBFO configurations with an active measurement study covering over 2.7 million domains from which about 2.1 million have MX records. Of these MX-enabled domains, about 60\% are observed to use BLBFO, and MX-based balancing seems more common than round-robin DNS. Email hosting services offer one explanation for this adoption rate. Many domains seem to also prefer fine-tuned configurations instead of relying on randomization assumptions. Furthermore, about 27\% of the domains have at least one exchanger with a valid IPv6 address. Finally, some misconfigurations and related oddities are visible.
\end{abstract}


\section{Introduction}

\thispagestyle{firststyle} 

Electronic mail uses the DNS to determine the Internet protocol (IP) addresses of the receiving mail servers. In other words, email, IP, and DNS establish one of the Internet's core functionalities. The standards for these protocols were also specified around the same time; the 1982 standard~\cite{RFC821} for email only slightly precedes the early DNS standards. What is more: already the later 1989 standard~\cite{RFC1123} specified also the two basic mechanisms for reliability of email delivery: (a) multiple MX records with preference values and (b) ``multi-homing'' with multiple IP addresses. The latter is nowadays closely tied to round-robin DNS. Regardless of the particular terminology used, both mechanisms are still used today for distributing network load and handling of mail delivery failures. However, no notable previous Internet measurement research appears to exist regarding the prevalence of these setups and their typical configurations. This gap in the literature provides the paper's motivation---and patching the gap provides the contribution.

\IEEEpubidadjcol

It must be emphasized that the paper's focus is also strictly restricted to these two \textit{basic} load-balancing and fail-over configurations. Many alternatives have been developed and deployed over the years. These solutions often extend particularly the load-balancing question toward more fine-grained hardware and software aspects~\cite{MoonKim05}. Also more fundamental infrastructural changes have occurred. A~good example would be the so-called split-horizon setups through which an optimal mail server is picked according to a client's IP address, which may be mapped to a specific network infrastructure or a particular geographic location~\cite{Aitchison05}. Content delivery networks (CDNs) are the prime example in this regard~\cite{PanHou03}. While CDNs are nowadays extensively used particularly for web and multimedia content, the BLBFO configurations are still frequently used for email delivery---as will be shown.

\section{Background}\label{sec: background}

\subsection{Fundamentals}\label{subsec: fundamentals}

\begin{figure}[th!b]
\centering
\includegraphics[width=8cm, height=3cm]{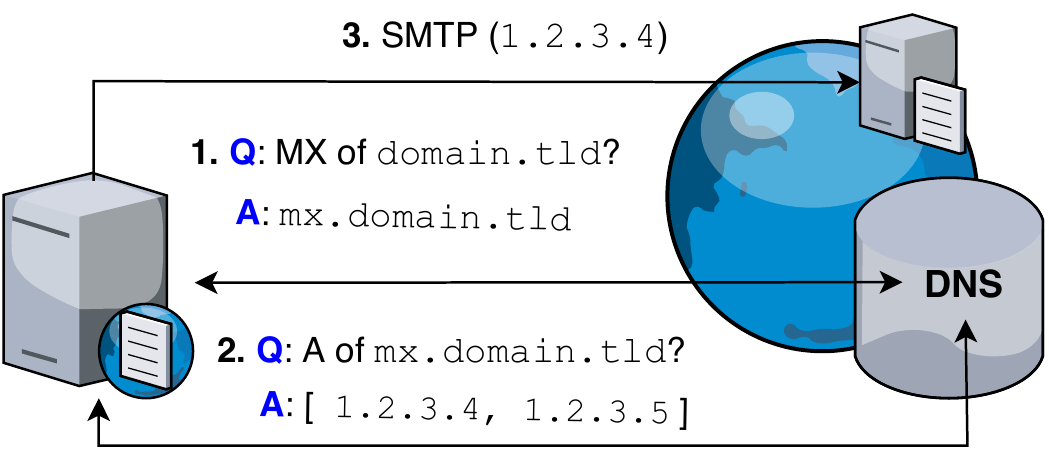}
\caption{Basic DNS Resolving for Email Delivery}
\label{fig: basic}
\end{figure}

The fundamental DNS aspects have remained surprisingly stable over the decades for email delivery. In essence: after having lexically identified a domain to which a mail will be delivered, a client's mail transfer agent (MTA) queries DNS to obtain the domain's mail exchanger resource records. These resource records specify the mail servers responsible for accepting emails on behalf of the domain. The records contain the fully qualified domain names (FQDNs) of the mail servers, the usual time-to-live (TTL) values for the records, and specific integer-valued preference or priority values that specify the domain's preferred FQDNs for the delivery; lower values are preferred over higher values. After the MTA has picked the FQDN preferred, it queries the A (IPv4) or AAAA (IPv6) records of the FQDN chosen in order to obtain the addresses to which a transmission control protocol connection is established via the simple mail transfer protocol (SMTP). A simple resolving scenario for delivery is illustrated in Fig.~\ref{fig: basic}.

To fix the notation, let $\vcm$, $\vcp$, $\vca$, and $\vcaaaa$ denote vectors with lengths $n_m$, $n_p$, $n_a$, and $\na$. Assume that $\vcm$ contains the FQDNs from the MX records of a given domain, $\vcp$ the preference values specified in the MX records, $\vca$ the A records of these FQDNs, and, finally, $\vcaaaa$ the potential AAAA records of the specified mail servers. An equality $n_m = n_p$ always~holds.

An inequality $n_m > 0$ is also assumed to hold. This assumption, however, does not mean that a given domain would not accept emails. If $n_m = 0$, meaning that no MX records were returned, the client's MTA assumes that the domain operates with so-called implicit MX records~\cite{RFC5321}. In this case it attempts to deliver the mail to the addresses of the domain's A or AAAA records (instead of the A or AAAA records of the FQDNs specified in the MX records). Furthermore, a domain may not accept emails even though $n_m > 0$ because so-called ``Null MX'' records may be used to prevent unnecessary delivery attempts~\cite{RFC7505}. After excluding these cases, it is assumed in the measurements that each domain name in $\vcm$ is a FQDN. In addition, each of these FQDNs is assumed to resolve to one or more IPv4 or IPV6 addresses. By implication, either $n_a > 0$, $\na > 0$, or both are non-zero---yet $n_m$ may not necessarily equal $n_a$~or~$\na$.

In reality, many other assumptions apply and a vast amount of additional checks are typically done before a client's email reaches its target. These assertions involve also DNS. To prevent spam, many email servers have long rejected emails from MTAs whose IPs do not have valid domain name pointer (PTR) records. In other words, a client who queries for MX and A (or AAAA) records without PTR records is usually up to no good~\cite{Whyte06}. The reverse also applies: email servers should also have valid PTR records, meaning that any given $a_i$ in $\vca$ or $\aaaa_j$ in $\vcaaaa$ should pass a reverse DNS lookup. Although the associated PTR records do not have to be forward-confirmed (that is, a PTR record of a $a_i$ or a $\aaaa_j$ points back to the given FQDN in a MX record), one-to-one mappings are generally recommended~\cite{RFC1912}. For instance, both \texttt{1.2.3.4} and \texttt{1.2.3.5} in Fig.~\ref{fig: basic} should thus have PTR records pointing to \texttt{mx.domain.tld}. It should be also stressed that aliases (CNAMEs) are prohibited for MX records~\cite{RFC821, RFC5321}. With respect to the running example, \texttt{domain.tld} may be a CNAME, but the mail exchanger \texttt{mx.domain.tld} may not.

A further point worth briefly remarking is the use of text (TXT) records to specify a sender policy framework (SPF) for hosts who are allowed to send emails on behalf of a domain~\cite{RFC7208}.  For instance, a TXT record with a value \texttt{"v=spf1~-all"}  announces that the given domain does not send mails. Related to these are the DMARC (domain-based message authentication, reporting and conformance) and DKIM (DomainKeys identified mail) standards, which both use also the DNS. According to recent Internet measurement studies, these specifications are frequently used nowadays; though, DKIM and SPF more often than DMARC~\cite{OTA19, Portier19}.

\subsection{Basic Load-Balancing and Fail-Over via DNS}\label{subsec: BLBFO}

\begin{figure}[t!]
\centering
\includegraphics[width=8cm, height=3.8cm]{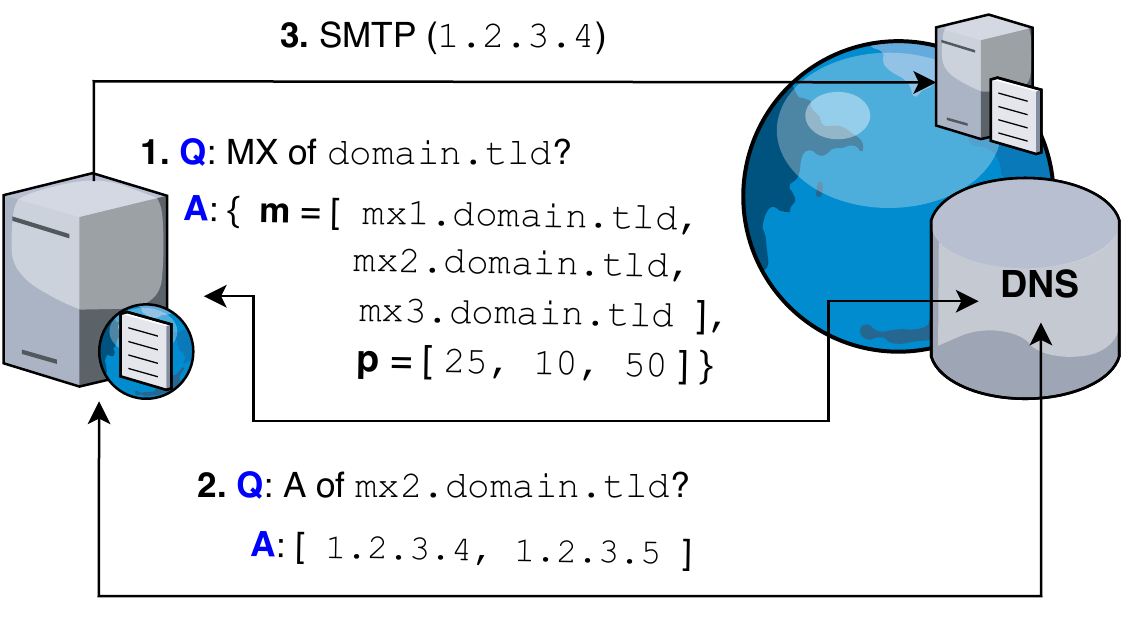}
\caption{Simple IPv4 Hybrid Balancing}
\label{fig: hybrid}
\end{figure}

There are two classical BLBFO solutions for email delivery. Neither one is accompanied with formal standards or rigorous specifications for well-defined behavior. Given that also the terminology is lax, the solutions can be labeled as (a)~\text{\textit{MX-balancing}} and (b)~\textit{round-robin DNS}. In addition, complex setups may use a (c)~``\textit{hybrid} strategy'' that combines the two. An example of a hybrid setup is shown in Fig.~\ref{fig: hybrid}.

In essence, MX-balancing specifies multiple MX records and uses the preference values for the BLBFO. In terms of fail-over, a typical setup contains one or few backup servers for which high preferences values are used to ensure delivery in times of high load. For instance, in Fig.~\ref{fig: hybrid} a client's MTA should prefer \texttt{mx2.domain.tld}, but if the two servers at \texttt{1.2.3.4} and \texttt{1.2.3.5} are busy, \texttt{mx1.domain.tld} would be next in the line. In terms of load-balancing, a classical option is to specify multiple MX records with the same preference value. In this case a client's MTA should pick a mail server randomly: when ``there are multiple destinations with the same preference and there is no clear reason to favor one (e.g., by recognition of an easily reached address), then the sender-SMTP MUST randomize them to spread the load across multiple mail exchangers for a specific organization''~\cite{RFC5321}. Most MTAs honor this mandate, although some allow to optionally alter the default randomization behavior~\cite{Aitchison05}.

Round-robin DNS is the classical alternative to MX-based balancing. The setup is typically implemented by using local replicas; a query is answered with a permuted list of records under the assumption that a client picks the first address from the list~\cite{PanHou03}. With this assumption, load-balancing occurs due to the permutation of the returned addresses at the next query. For instance, in both Fig.~\ref{fig: basic} and Fig.~\ref{fig: hybrid} two IPv4 addresses are returned for the second DNS query, but the client's MTA initiates connection to the first of these via SMTP. In addition to permutation, popular DNS servers allow to configure also randomization and fixed ordering~\cite{Aitchison05}. Further complexity is added by client applications, which have the final say in picking their preferred addresses. Although most current applications likely conform with the standards~\cite{RFC6724}, using the first address delivered via a \texttt{getaddrinfo} system call and then moving to the next one in case of a failure, there are no guarantees that all applications follow this behavior. According to measurements, the majority of client applications indeed pick the first address, although some seem to choose also randomly~\cite{Callahan13}. Besides these assumptions, traditional, \text{on-site}, round-robin DNS has become less relevant particularly for A records due to the global adoption boom of CDNs.

Both MX-based balancing and round-robin DNS contain also other obvious limitations. For instance, round-robin DNS requires that each replica used for the balancing is IP-addressable~\cite{PanHou03}. A more fundamental issue relates to TTL values and caching. The issue is usually framed with a trade-off: specifying a TTL value close to zero increases the effectiveness of balancing but decreases caching and thus increases also the load. While the controversial question about appropriate TTLs has long been debated, recent measurements indicate a trend toward low TTL values~\cite{APNIC19a, Moura19}. The explanation largely again traces to content delivery networks.

\section{Related Work}

The domain name system has been extensively studied and measured in recent years. There have been several large-scale data collection frameworks for passively measuring DNS traffic from different vantage points~\cite{Callahan13, Foremski19}. However, the present work belongs to the category of active measurements, which essentially query DNS to obtain information about a predefined set of domains. Within this active measurement domain, the questions examined typically focus on some particular resource records. Examples include A, AAAA, CAA, CNAME, NS, and SOA records. Adoption of standards~\cite{Ruohonen19JCST}, security~\cite{Kontouras16}, and misconfigurations~\cite{LuDong14} have provided the typical motivations for these record-specific studies.  Some of these have touched also email-specific DNS records. For instance, SPF configurations and the free-form TXT records have been examined recently~\text{\cite{Portier19, Scheffler18}}. Although also MX records have been measured~\cite{LuDong14}, a reasonably comprehensive literature search indicates no directly related previous works regarding the BLBFO theme. Likewise, there are studies on round-robin DNS~\cite{PanHou03, Cheung03}, but limited
previous work exists regarding its deployment and use in the email delivery context. The gaps in the literature is noteworthy and surprising because the BLBFO setups considered are classical and often encountered by network administrators during DNS~configuration.

\section{Data and Measurements}\label{sec: data}

The initial dataset is based on Alexa's top-million (1M) domain name popularity lists. Although these lists are frequently used for different measurements, the lists carry many well-known limitations that should be taken into account before blindly using the lists. Three such limitations can be briefly remarked. First, the lists are biased toward large organizations who host their popular domains on CDNs and related large-scale network infrastructures~\text{\cite{Ruohonen19JCST, Zimermann17}}. Second, the lists contain considerable longitudinal variation particularly during weekends~\cite{Rweyemamu19}. Third, the lists are not curated for observing AAAA records and dual-stack deployments~\cite{Bajpai15}. Although different between-list merge solutions and practical recommendations have been recently proposed~\cite{Rweyemamu19, LePochat19}, these have mostly focused on security research and web-specific measurement contexts. In other words, it remains unclear how the limitations affect MX-based mail delivery~measurements.

In this context, it is preferable to have a relatively large list instead focusing on a sharper set of particularly popular domains. The reason is simple: email delivery is not usually the reason why many domains are popular---in fact, some popular domains operate without MX records. To further account for the weekly variation, the initial dataset was assembled by including all unique domain names present in Alexa's seven individual 1M lists that were available from a repository maintained by the Technical University of Munich~\cite{TUM19}. These lists cover a whole week between the 4th and 10th of November 2019. Although the popularity ranks should be approached with caution~\cite{Rweyemamu19}, median was used across all ranks reported for a domain in the weekly 1M lists. The domain names were not manipulated; domains as well as their subdomains are covered. In total, \text{$k_q = 2,709,827$} domains were resolved.

The actual resolving was done in three steps via live DNS using Google's name server at \texttt{8.8.8.8}. Although multiple passes are sometimes carried out in order to account for timeouts and related errors~\cite{Ruohonen19JCST, Ruohonen17CIT}, each resolving step was implemented in a single pass. Thus, in the first step MX and TXT records were queried for each domain. If MX records were not found for a given domain, the domain was excluded from the sample analyzed. The same applies to errors, whether NXDOMAIN cases or timeouts. In the second step both the A and AAAA records were resolved for each MX record of each domain. In the third and final step the PTR records of the A and AAAA records of the MX records were queried.

Finally, it is necessary to point out that no duplicate queries were made. Although TTLs are difficult to evaluate on the client-side~\cite{Moura19}, this uniqueness of the queries should ensure that the TTL values approximate the typical values supplied by the given resolver for MX records in Northern Europe.

\section{Results}\label{sec: results}

\subsection{Sample Characteristics}

Only $k_w = 646,339$ domains from the approximately 2.7M domains queried did not have a single MX record. In other words, as much as $76\%$ of the popular domains announced capability for email delivery. Thus, the sample analyzed covers \text{$k = k_q - (k_w  + 269)$} domains. The additional $269$ domains excluded refer to the ``Null MX'' setups. To use the notation from Subsection~\ref{subsec: fundamentals}, these setups were identified by checking that $n_m = 1$, and then verifying that the single $p_1$ present in $\vcp$ equaled zero and $m_1$ in $\vcm$ equaled a single dot character~\cite{RFC7505}.

\subsection{Record Counts}\label{subsec: record counts}

\begin{figure}[t!]
\centering
\includegraphics[width=\linewidth, height=3.5cm]{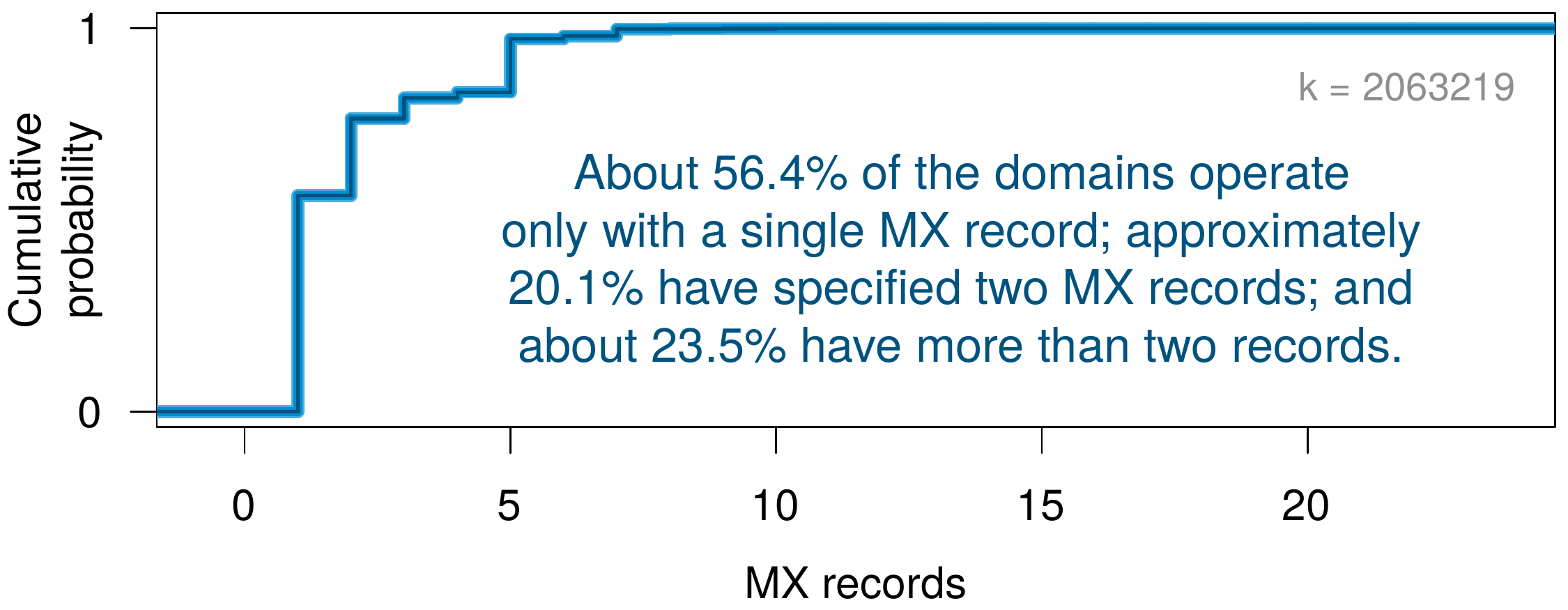}
\caption{Number of MX records}
\label{fig: mx n}
\end{figure}

The MX record counts provide a good way to start the empirical analysis. These are thus shown in Fig.~\ref{fig: mx n}. By a thin margin, domains with only a single MX record surpass those running with multiple records. In the latter group most of the domains run with $2-5$ mail exchangers---only about $2.7\%$ of the domains have specified more than five records. The maximum of twenty records was specified by only one domain.

\begin{figure}[t!]
\centering
\includegraphics[width=\linewidth, height=4cm]{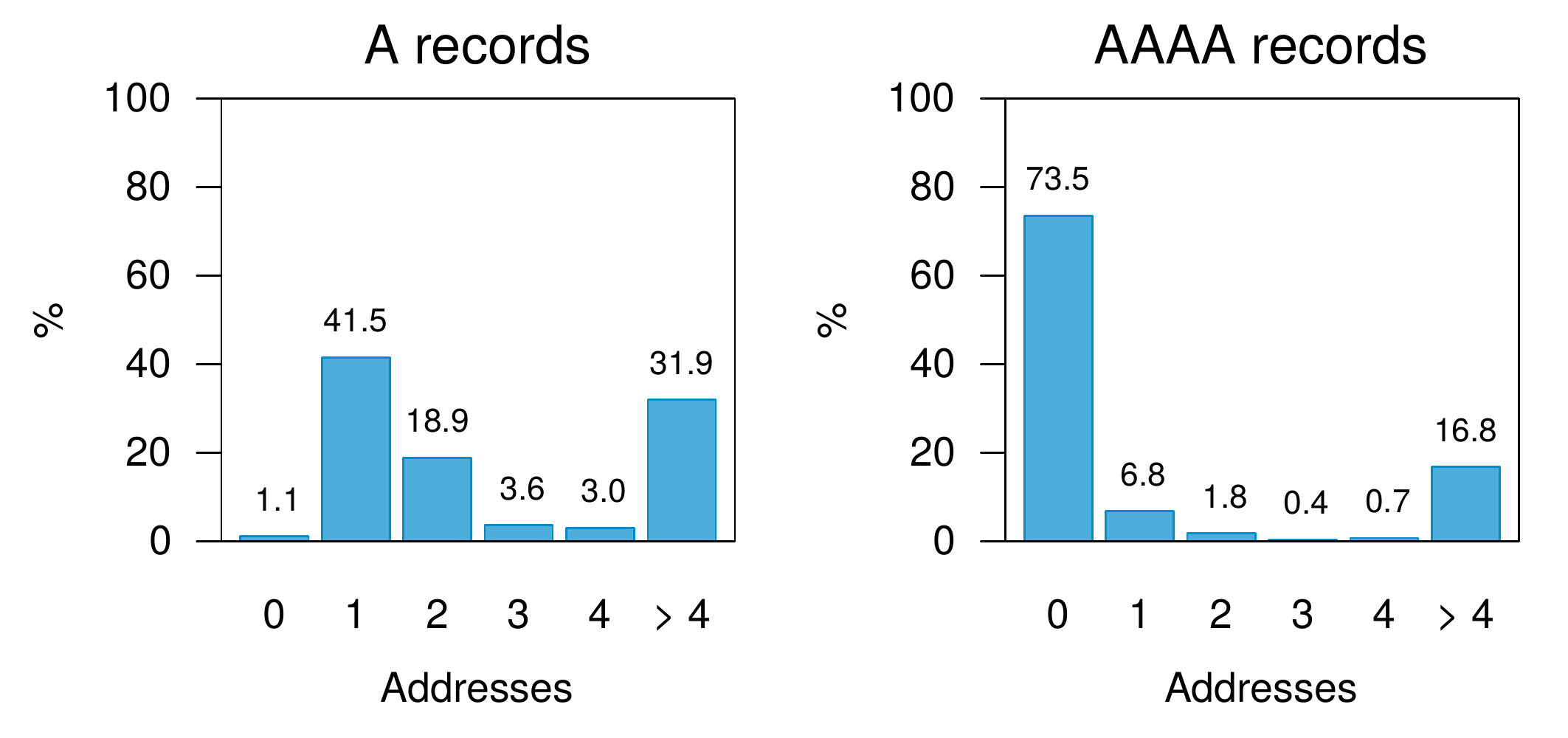}
\caption{Number of A and AAAA records}
\label{fig: addresses}
\end{figure}

Turning to the IP addresses of the MX records, the relative share of A and AAAAA records is summarized in Fig.~\ref{fig: addresses}. Five brief points can be enumerated about these IP address counts.

\begin{enumerate}
\item{Even though approximately $41.5\%$ of the domains operate with just a single A record, there are domains whose mail exchangers resolve to even up to fifty IPv4 addresses. There is also one extreme outlier: \texttt{isllc.com}, which has specified $17$ MX records (all subdomains of \texttt{google.com} and \texttt{barracudanetworks.com}). Together these exchangers resolved to $173$ A records.}
\item{IPv6 adoption is---even a little unexpectedly--- relatively strong: about $26.5\%$ of the domains have at least one exchanger that resolves to at least one IPv6 address. This amount is in line with IPv6 adoption trackers~\cite{APNIC19b}.}
\item{The MX record counts in Fig.~\ref{fig: mx n} are correlated with the address counts: the Pearson's product-moment correlation coefficients are $0.30$ and $0.71$ for the MX and A, and MX and AAAA record counts, respectively. These correlations hint about the presence of hybrid setups.}
\item{Not all of the IPv4 and IPv6 addresses are unique; some of the unique MX records of some particular domains resolve to the same addresses. It is difficult to say anything definite about this use of non-unique addresses; email hosting services, CDNs, and configuration mistakes each offer a slightly different but plausible~explanation.}
\item{About $1.3\%$ of the domains in the sample have specified at least one MX record that does not resolve to a valid IPv4. To rule out data collection issues such as timeouts, these cases were re-checked; all truly are NXDOMAIN cases. While IPv6-only domains are possible, so are configuration mistakes and network maintenance issues.}
\end{enumerate}

\subsection{Configurations}\label{subsec: configurations}

As was noted in Subsection~\ref{subsec: BLBFO}, there exists neither a well-established terminology nor a ready-made topology for measuring BLBFO setups. What can be defined unambiguously, however, are the following simple, non-BLBFO setups:

\begin{itemize}
\item{\textit{Plain IPv4-only}: $n_m = 1\land n_a = 1\land\na = 0$. That is, only a single MX record is present, and the FQDN specified in the resource record resolves to a single IPv4 address.}
\item{\textit{Plain IPv6-only}: $n_m = 1\land n_a = 0\land\na = 1$. A case in which delivery occurs only through a single IPv6 address.}
\item{\textit{Plain dual-stack}: $n_m = n_a = \na = 1$, that is, only a single server is specified in the MX-FQDN vector $\vcm$ but it serves mail transfer agents through both IPv4 and IPv6.}
\end{itemize}

Analogous but less robust definitions can be used for the BLBFO configurations. To simplify the empirical analysis, the following definitions omit the IP version categorization:

\begin{itemize}
\item{\textit{Round-robin}: $n_m = 1 \land (n_a > 1 \lor \na > 1)$. In other words, a domain uses a single MX record, which is mapped either to multiple IPv4s or multiple IPv6s. A simple $n_a = 2$ round-robin setup is present in Fig.~\ref{fig: basic}.}
\item{\textit{MX-balancing}: $n_m > 1 \land (n_a = n_m \lor \na = n_m)$.}
\item{\textit{Hybrid}: $n_m > 1$, and the given domain has not already been classified as operating with the noted MX-balancing.}
\end{itemize}

Here, a conceptual problem relates particularly to the demarcation between MX-balancing and hybrid setups. Nevertheless, the definitions provide a decent glimpse on different BLBFO setups even though the construct validity is not perfect. While keeping this point in mind, the results are shown in Table~\ref{tab: simple setups}.

\begin{table}[th!b]
\centering
\caption{Simple and BLBFO Configurations}
\label{tab: simple setups}
\begin{tabular}{llrr}
\toprule
Configuration & Type & Frequency & Share ($\simeq$ \%) \\
\hline
Simple & Plain IPv4-only & $817,348$ & $39.6$ \\
& Plain IPv6-only & $48$ & $<0.01$ \\
& Plain dual-stack & $34,038$ & $1.6$ \\
\cmidrule{2-4}
BLBFO & Round-robin & $291,325$ & $14.1$ \\
& MX-balancing & $632,994$ & $30.7$ \\
& Hybrid setup & $266,673$ & $12.9$ \\
\cmidrule{2-4}
Others & Non-identified & $20,793$ & $1.0$ \\
\bottomrule
\end{tabular}
\end{table}

About 41.3\% of all domains observed operate with simple setups without neither basic DNS-based load-balancing nor fail-over. Roughly about three-fifths use BLBFO configurations; MX-balancing is more common than classical round-robin DNS specified with a single MX record and multiple addresses. Different hybrid setups are also relatively common.

\subsection{Popularity}

\begin{figure}[th!b]
\centering
\includegraphics[width=\linewidth, height=3cm]{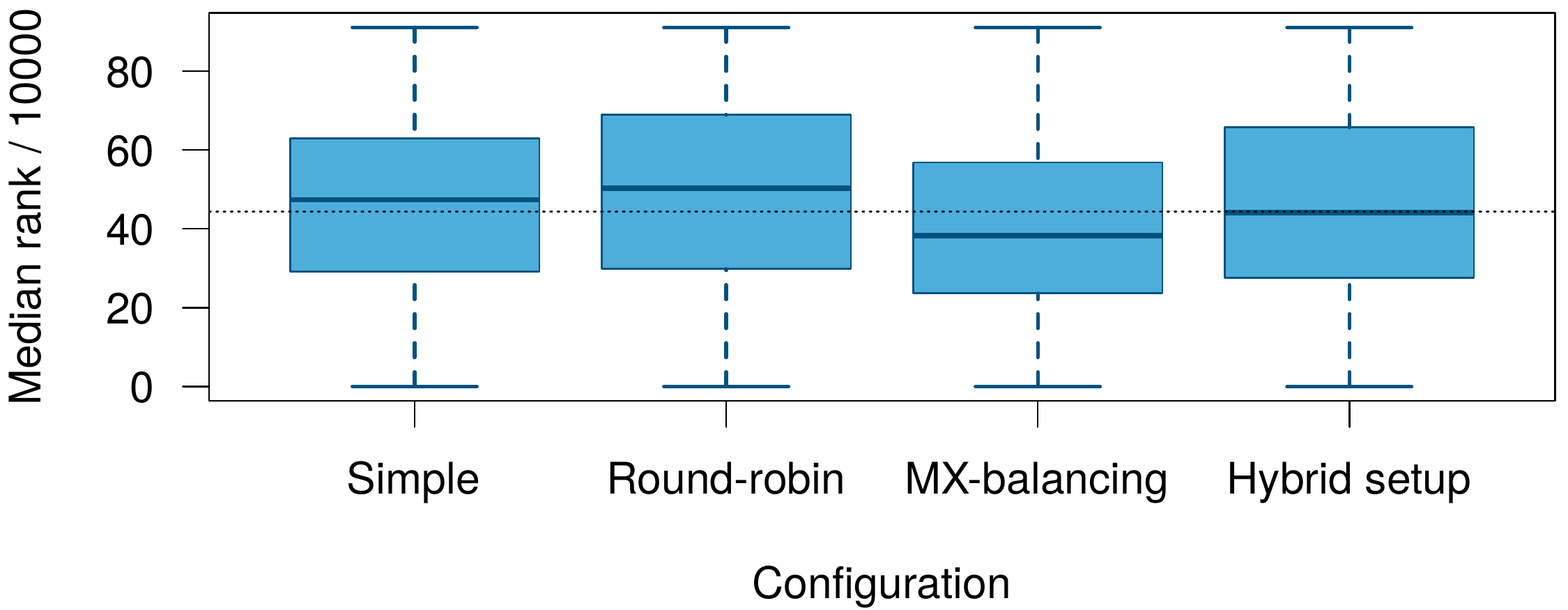}
\caption{Median Alexa's popularity ranks across configurations}
\label{fig: ranks}
\end{figure}

The median popularity ranks of the weekly Alexa 1M lists provide a decent metric to probe whether particularly popular domains are more likely to use BLBFO configurations. While these ranks should be interpreted with care, the topology used is coarse enough to avoid the spurious correlations often seen for individual (domain-level) analysis~\cite{Rweyemamu19}. Thus, according to Fig.~\ref{fig: ranks}, the medians of the median weekly ranks are quite similar across the three simple and the three BLBFO setups. Though---partially due to the large sample size, the differences are statistically significant according to the Kruskal-Wallis~test.

\subsection{Hosting}

To probe the use of email hosting services, Table~\ref{tab: hosting} shows a cross-tabulation of the configurations across two popular hosting services. These are defined according to whether at least one MX record referred to the subdomains of \texttt{outlook.com} (Microsoft) and \texttt{google.com} or \texttt{googlemail.com} (Google). About 24\% of the domains have specified at least one MX record pointing to these two email services. As could be expected, only a relatively few of the domains operating with simple setups use these services.

\begin{table}[t!]
\centering
\caption{Configurations and Two Popular Email Hosting Services (\%)}
\label{tab: hosting}
\begin{tabular}{lcccc}
\toprule
& Simple & Round-robin & MX-balancing & Hybrid setup \\
\hline
Hosting & $2.1$ & $41.5$ & $52.2$ & $8.6$ \\
Others & $97.9$ & $58.5$ & $47.8$ & $91.4$ \\
\bottomrule
\end{tabular}
\end{table}

\subsection{Preference Values and TTLs}

\begin{figure}[th!b]
\centering
\includegraphics[width=\linewidth, height=2.5cm]{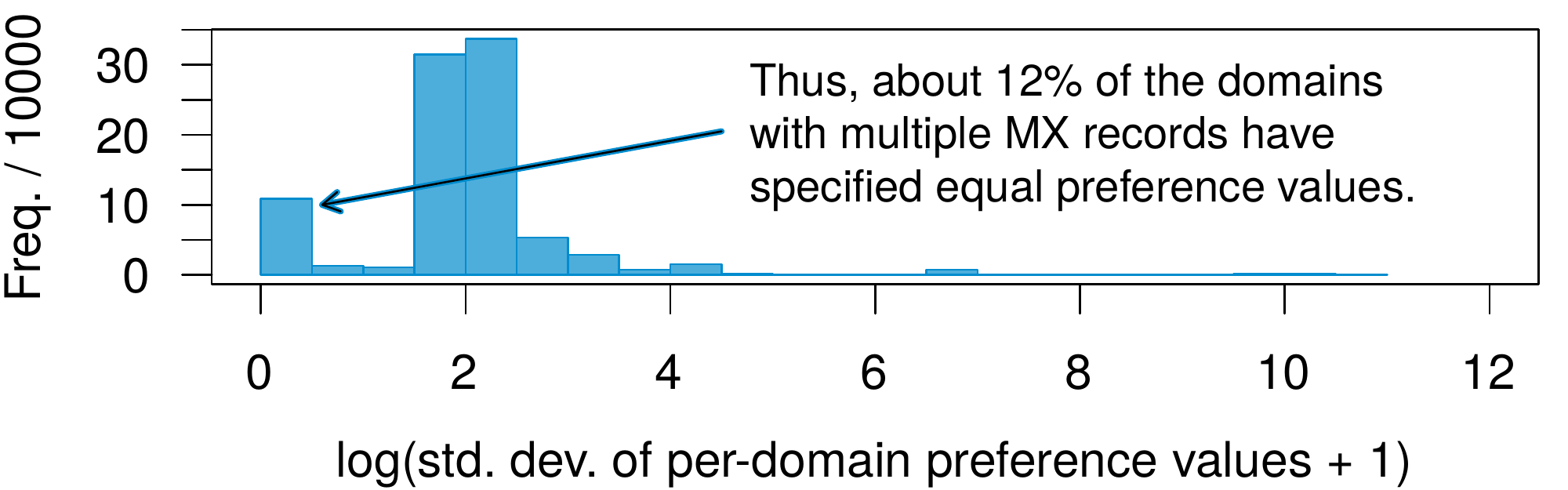}
\caption{Distribution of the standard deviation of the preference values}
\label{fig: preference values}
\end{figure}

The preference values for MX records are specific to a given configuration; therefore, the average values for these do not provide any reliable Internet-wide insights. However, the variance of these, among other things, reveals the prevalence of preferences for random client-side picking of the mail exchangers (cf.~Subsection~\ref{subsec: BLBFO}). According to Fig.~\ref{fig: preference values}, which shows a histogram of the standard deviation of each $\vcp$ in a subset of domains with $n_m > 1$, about 12\% of the domains have specified equal preference values. Although it is difficult to say whether this is a low or a high amount, it seems fair to remark that most of the MX-based and hybrid setups observed are fine-tuned for BLBFO instead of relying on randomization.

\begin{figure}[th!b]
\centering
\includegraphics[width=\linewidth, height=2.5cm]{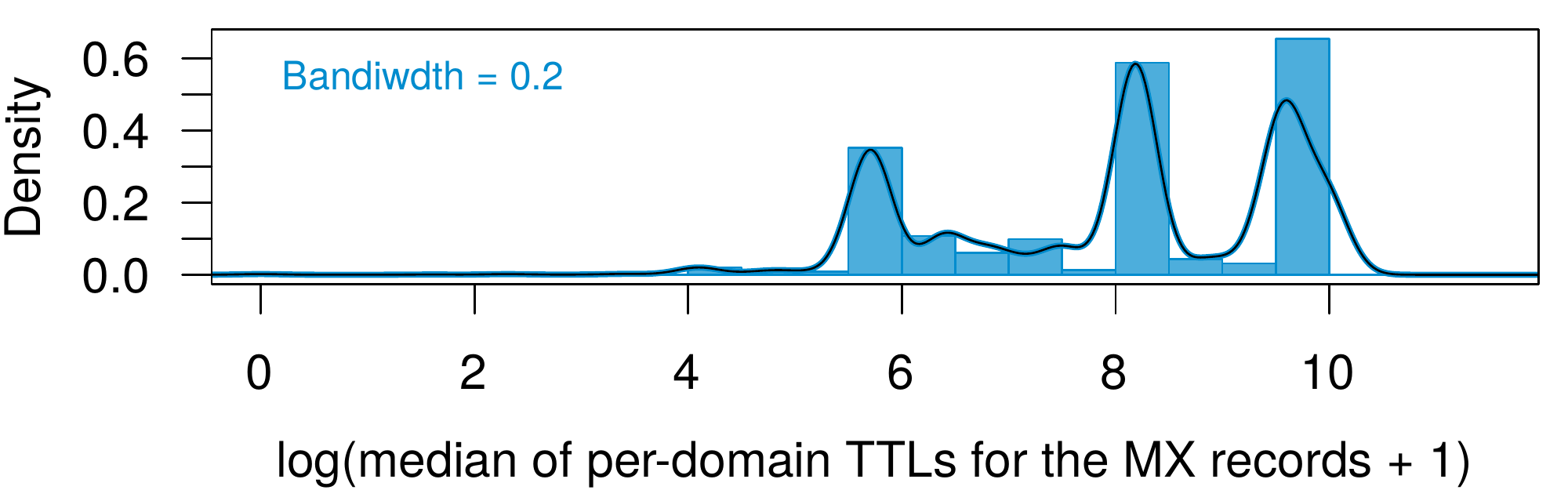}
\caption{Distribution of the TTLs of the MX records}
\label{fig: ttls}
\end{figure}

A further point can be made about the TTL values shown in Fig.~\ref{fig: ttls}. When compared to previous measurements \cite{Callahan13, Moura19}, most of these values are relatively large. For instance, only about one-third of these values are below $2000$. Although the reliability of the TTL values is unclear due to the active measurement approach, it seems sensible to tentatively conclude that higher values are used for MX records than for A records.

\subsection{Misconfigurations and Outliers}

In addition to the non-unique IPs and the \text{NXDOMAIN} cases already noted in Subsection~\ref{subsec: record counts}, three points can be made about potential misconfigurations and other details. First, the bright things: as much as 68\% of the domains have specified a SPF entry; these were probed simply by searching the \texttt{v=spf} character string from each TXT record. This share is comparable to previous measurements~\cite{Portier19}, and email hosting services are one explanation for the increased adoption~\cite{Durumeric15b}. Though, interestingly, there are no differences between the configurations in this regard; both simple (69\%) and BLBFO (68\%) setups tend to use SPF. Second, some blatant issues are present: $970$ and $125$ domains have specified at least one A or AAAA record, respectively, pointing to a private network address or a local host (as defined in \cite{RFC1918} and \cite{RFC4193}). Although the reasons for these outliers are not well-known, a similar observation has been made also previously~\cite{LuDong14, Hendriks17, Plonka08}. Third, the absence of PTR records for the A and AAAA records of the MX records is relatively common: about 16\% and 10\% of the domains have at least one IPv4 or IPv6, respectively, for which a PTR lookup fails. This observation correlates with the NXDOMAIN cases. Also this lack of PTRs has been observed in previous Internet~measurements~\cite{LuDong14}.

\section{Limitations}

Four limitations should be briefly noted. The first relates to the conceptual problems (see Subsection~\ref{subsec: configurations}). In other words, it is not entirely clear how the three configuration types should be defined and measured. By implication, the results are more robust when the BLBFO setups as a whole are compared to simple setups. The second problem is closely related: as with CDNs and other large-scale network infrastructures~\cite{Ruohonen17CIT}, particularly the IP addresses returned for email hosting services may evolve over time. For instance, a query could return two addresses as in Figs.~\ref{fig: basic}~and~\ref{fig: hybrid}, whereas another query might return a single address or three addresses. By implication, the results regarding round-robin DNS may not be entirely reliable across all domains. A longitudinal measurement framework is required for examining this assumption. The third limitation is about the same theme: only a single client-side vantage point was used for the measurements. In addition to longitudinal approaches, further work should focus on measurements from multiple vantage points. Distributed frameworks (such as RIPE Atlas) seem prolific in this regard. Finally, the last limitation is fundamental: it seems difficult---if not impossible---to deduce about the actual number of mail servers via DNS alone. Multiple servers may be behind a single IP address, and so~on.

\section{Conclusion}\label{sec: conclusion}

This paper surveyed the use and configuration of basic DNS-based load-balancing and fail-over setups for email delivery. To summarize the results: BLBFO setups are common; MX-based balancing seems more common than round-robin DNS; fine-tuning instead of randomization is a preferred strategy for the MX-based setups; email hosting services have pushed the adoption of BLBFO in general; and, finally, some configuration mistakes and other peculiarities are present.

\balance
\bibliographystyle{IEEEtran}


\end{document}